# Quantum superballistic diffusion in presence of white noise

$E. Gholami^{1*}$  $Z. Mohammaddost\ Lashkami^{2}$

[1] Department of Physics, Isfahan University of Technology, Isfahan 8415683111, Iran
[2] Department of Physics, Zanjan University, 45196, Zanjan, Iran

As an unusual type of anomalous diffusion behavior, (transient) superballistic transport is not well understood but it has been experimentally observed recently. We here calculate the white noise effect (in Markov approximation) on the quantum diffusion in 1D tight-binding model with disordered and quasi-periodic region of size L attached to perfect lattices at both ends. Here we show the effect of white noise on spreading of the wave packet initially located at the center of the sublattice. We find threshold values of the white noise strength, beyond which the quantum hyperdiffusion does not occur. We predict from our numerical studies that the quantum hyperdiffusion exponent can be tuned by the strength of applied white noise. Therefore we can manually apply noise to a system to drive it to desired diffusion rate.

The quantum diffusion in 1D tight-binding model has an enriched background [1] [2] [3]. First numerical evidence supporting that the variance of a wave packet in 1D tight-binding can show a superballistic increase $M(t) = t^v$ with $2 < v \leq 3$ for parametrically large time intervals and the appropriate model was constructed by L. Hufnagel et al [4]. The model explains this phenomenon and its predictions were verified numerically for various disordered and quasiperiodic systems. They replaced the disordered part by a point source and anything emitted from it moves with a constant velocity v modeling the dynamics of a perfect lattice [4]. Then superballistic diffusion of entanglement in disordered spin chains was constructed [5]. In 2012 Z. Zhang et al. showed a superballistic increase $M(t) = t^v$ with $3 < v \leq 4.7$ and they extended the interpretation given in Ref. [4] to the nonlinear regime [6]. It was in 2013 that the superballistic growth of the variance have seen experimentaly for optical wave packets. [7]

The fractal [1] [8] [9] and multifractal [10] analyze of the width of a spreading wave packet reveals that for systems where the shape of the wave packet is preserved, the kth moment evolves as $t^{k\beta}$ with $\beta = D_2^\mu / D_2^\psi$, while, in general, $t^{k\beta}$ is an optimal lower bound. $D_2^\mu$ is the correlation dimension of the spectral measure μ (the local density of states) and $D_2^\psi$ is the correlation dimension of the (suitably averaged) eigenfunctions.

The (disorder or phase-averaged) diffusion exponent is of particular physical importance because it characterizes the low-temperature behavior of the direct conductivity as given by Kubo's formula in the relaxation time approximation. [11]

Throughout this article we described the system in universal terms, not specific to matter waves, as manifested by the analogy between the Schrödinger equation and the paraxial wave equation. Hyper-diffusion is in fact a universal concept, which should be observable in a variety of systems beyond matter waves, such as optics, sound waves, plasma, and in the transport of conduction electrons in semiconductors [12]. Furthermore, fundamentally, once such temporal acceleration would reach very high velocities, relativistic effects would have to be included. Most certainly, these ideas open a range of exciting possibilities. However, in view of the recent experiments on quantum walks of correlated photons [13] and on localization with entangled photons [14], it would be extremely interesting to know whether the phenomenon of hyper-transport would occur also with entangled photons as it occurs on entangled spin chains [5].

With regard to the decoherence problem, the temperature effect on wave packet spreading is an essential feature [15]. The theoretical description of relaxation and decoherence processes in open quantum systems often leads to a non-Markovian dynamics which is determined by pronounced memory effects. Strong system-environment couplings, correlations and entanglement in the initial state, interactions with environments at low temperatures and with spin baths, finite reservoirs, and transport processes in nanostructures can lead to long memory times and to a failure of the Markovian approximation [16]. But since here in calculating white noise effect on super-ballistic diffusion in 1D tight binding lattice, we do not deal with these restrictions, we can use the Markov approximation and the Lindblad equation. In Strong system-environment couplings we can also use a similar method called Non-Markovian generalization of the Lindblad theory of open quantum systems [16] [17].

In this work we examine noise effect on quantum wave packet dynamics in several nonuniform 1D tight-binding lattices, where a sublattice with on-site potential is embedded in a lattice with uniform potential. Irrespective of whether the sublattice on-site potential is disordered or quasiperiodic (Some cases in absence of any environment where studied in Ref. [4], [6]). We find threshold values of the white noise strength, beyond which the quantum hyperdiffusion does not occur (in the disordered case, the observed disappearance of hyperdiffusion is based on a fixed number of realizations of the sublattice).

Such threshold values for disappearance of quantum hyperdiffusion should be one key element in real experimental studies, where the environment and noise have significant dephasing effect. Furthermore, we predict from our numerical studies that the quantum hyperdiffusion exponent can be extensively tuned by amount of induced white noise. That is we can manually induced noise to a system to drive it to an exact amount of diffusion rate. A phenomenological explanation for quantum hyperdiffusion exponents was discussed in Ref. [6]. The results must be within reach of today's cold-atom experiments and in case of presence of noise, it can also presence the lattices in solid state physics.

Consider the simple 1D tight-binding Hamiltonian:
$H = -\sum_{i,j}(t_{ij}c_i^+ c_j + t_{ij}^* c_j^+ c_i) + \sum_{i\in[-L,L]} V_i c_i^+ c_i$
Where $V_i$ represents the dimensionless on-site potential scaled by a tunneling rate, for a sublattice of length (2L+1) with $i \in [-L, L]$. That is, $V_i \neq 0$ only when $i \in [-L, L]$. The first term describes the hopping from site j to i and the second term describes the hopping from i to j. Because the Hamiltonian is Hermitian, the two coefficients must be equal to each other. In this study we only consider the case for first neighbors that is $t_{ij} \neq 0$ just for $i = j \pm 1$.

At time zero a localized wave packet is launched in the sublattice center, with $\rho_{n,m}(t=0) = \delta_{c,c}$, where "c" denotes the central cite in the lattice. This initial state is a coherent superposition of many quasimomentum eigenstates. We use the master equation of the Lindblad form that can be written in the form [18]
$\frac{\partial \rho}{\partial t} = L\rho \equiv -\frac{i}{\hbar}[H,\rho] + \sum_i \frac{1}{2}\gamma_i \mathcal{M}_i\{2A_i\rho A_i^+ - \rho A_i^+ A_i - A_i^+ A_i\rho\} + \sum_i \frac{1}{2}\gamma_i \aleph_i\{2A_i^+\rho A_i - \rho A_i A_i^+ - A_i A_i^+\rho\}$
more general noise term especially the local ones can found on Ref. [19] [20]. So the new Lindblad equation can be rewritten as:
$\frac{\partial \rho}{\partial t} = L\rho \equiv -\frac{i}{\hbar}[H,\rho] + \sum_i \Gamma_i\left\{A_i\rho A_i^+ - \frac{1}{2}\rho A_i^+ A_i - \frac{1}{2}A_i^+ A_i\rho\right\}$
where $\Gamma_i$ denotes noise intensity.

We assume that the interaction of the system with the environment is dominated by white-noise captured within the Haken-Strobl model (pure-dephasing) [21]. With the generators $A_i = |i\rangle\langle i|$. The dephasing term damps all off-diagonal entries of the density matrix, suppressing superpositions of localized states at a rate $\Gamma_i$, which is called the dephasing rate. Note that the pure-dephasing (Haken-Strobl) model is a simplified but useful model that has been successfully used in numerous studies in quantum optics, quantum information science, physical chemistry, and condensed matter physics. Its prediction becomes more realistic when the system is interacting with a thermal bath at high temperatures, where its effects can be modelled by white noise [22]. We measure the spreading of the wave packet by it's variance. The variance of the wave packet is defined as
$$variance \equiv \sigma^2 \equiv \sum_n n^2 |\psi_n(t)|^2$$
Where n is the lattice site index and $\psi_n(t)$ depicts a normalized time-evolving wave packet.

*Disordered case.* —We now consider a disordered sublattice; i.e., $V_i$ takes -V or +V randomly, for $i \in [-L, L]$, and $V_i = 0$ otherwise. Here we study the case for $\Gamma_i = 0, 0.01, 0.4, 0.1$ in fig. 1.

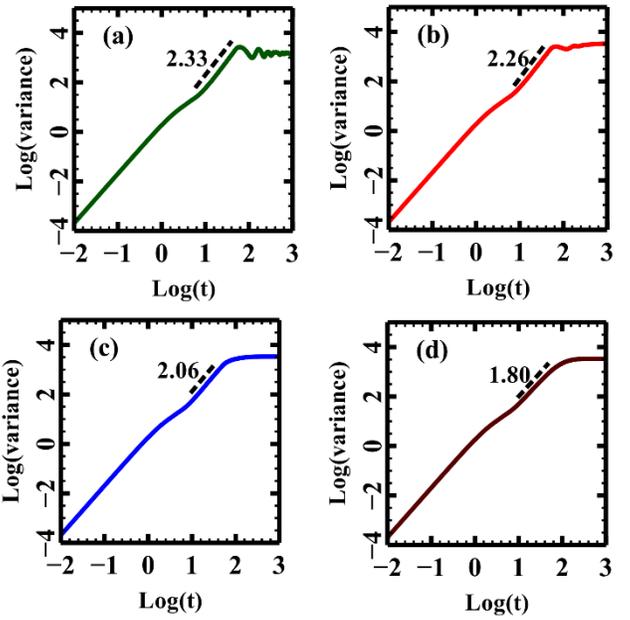

Figure 1: Time dependence of variance of the wave packet for a disordered sublattice (a) without any noise (b) for $\Gamma_i = 0.01$ (c) for $\Gamma_i = 0.04$ (d) for $\Gamma_i = 0.1$. Here and in all other figures, dashed lines represent a linear fit and all ploted quantities are dimensionless.

As we can see in fig. 1(a)-(b)-(c) the variance grows superballisticlly but it's power decreases with respect to strength of the induced noise, and in fig. 1(d) the induced noise finally causes the subballistic growth of the variance. So as we see, here the noise have negative effect on the spreading of the wave packet.

*Quasiperiodic (Harper) case.* — Quasiperiodic systems are between periodic and disordered systems. we now turn to a non-interacting Harper sublattice [23] with Aubry–André´ Hamiltonian [24] in a particular way, that is
$H = J\sum_i(|w_i\rangle\langle w_{i+1}| + |w_{i+1}\rangle\langle w_i|) + \Delta\sum_{i\in[-L,L]}\cos(2\pi\beta i + \phi)|w_i\rangle\langle w_i|$.

where $|w_i\rangle$ is the Wannier state localized at the lattice site i, J is the site-to-site tunneling energy(we chose $J = -1$, $\Delta$ is the strength of the disorder(we study $\Delta = 0.5, 1.5, 2.5$), $\beta = k_2/k_1$ is the ratio of the two lattice wave numbers, and $\phi$ is an arbitrary phase(we set $\phi = 0$). In the experiment, the two relevant energies J and $\Delta$ can be controlled independently by changing the heights of the primary and secondary lattice potentials, respectively. For a maximally incommensurate ratio $\beta = (\sqrt{5} - 1)/2$, the model exhibits a sharp transition from extended to localized states at $\Delta/J=2$ [24]. In fig. 2 we present the diffusion rate for Harper potential for $\Delta = 0.5$. In the absence of any noise in fig. 2(a) $\sigma^2 = t^{2.1}$ as the white noise become stronger fig. 2(c)-(d) the spreading of the wave packet becomes slower $\sigma^2 < t^2$.

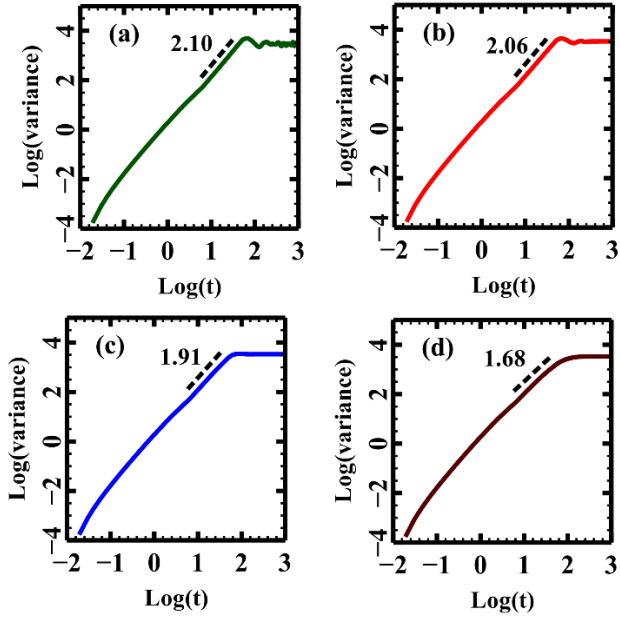

Figure 2: Time dependence of variance of the wave packet for a Harper sublattice for $\Delta = 0.5$. (a) without any noise (b) for $\Gamma_i = 0.01$ (c) for $\Gamma_i = 0.04$ (d) for $\Gamma_i = 0.1$.

In fig. 3 we present the diffusion rate for Harper potential for $\Delta = 1.5$. Here in fig. 3(b)-3(c) the white noise up to some threshold value improve the spreading of the wave packet and after it in fig. 3(d) the white noise begin to decrease the spreading of the wave packet.

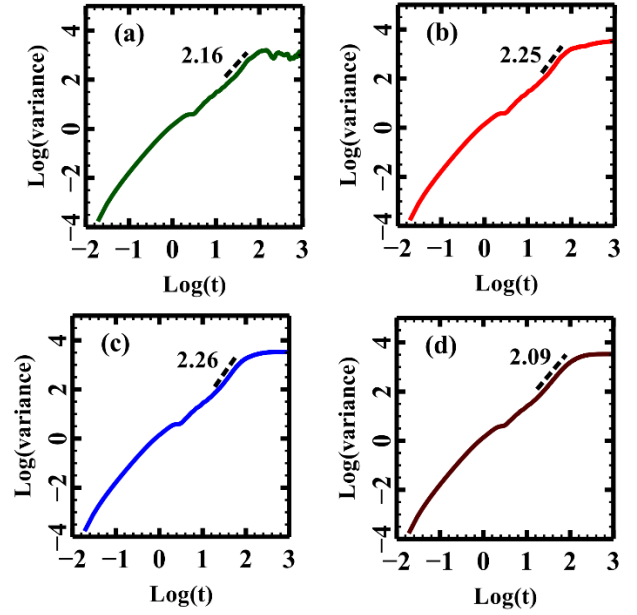

Figure 3: Time dependence of variance of the wave packet for a Harper sublattice for $\Delta = 1.5$. (a) without any noise (b) for $\Gamma_i = 0.01$ (c) for $\Gamma_i = 0.04$ (d) for $\Gamma_i = 0.1$.

In fig. 4 we present the diffusion rate for Harper potential for $\Delta = 2.5$.

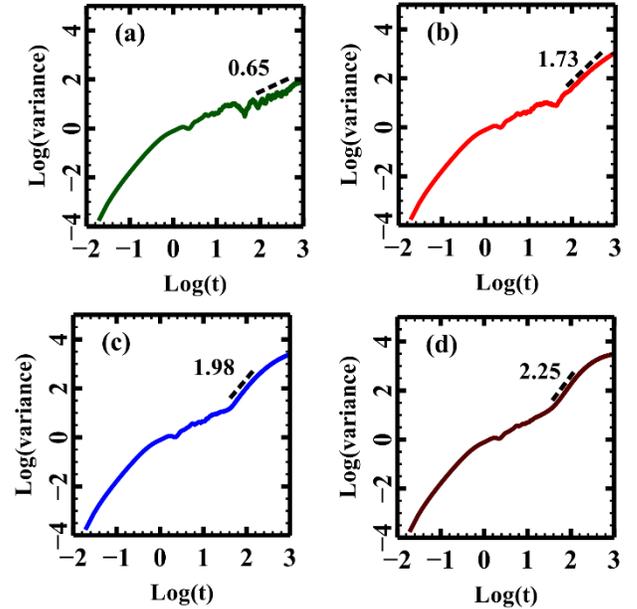

Figure 4: Time dependence of variance of the wave packet for a Harper sublattice for $\Delta = 2.5$. (a) without any noise (b) for $\Gamma_i = 0.01$ (c) for $\Gamma_i = 0.04$ (d) for $\Gamma_i = 0.1$.

Here noise not only improve spreading of the wave packet but also change the diffusion regime once from subdiffusion $0 < \nu \leq 1$ in fig. 4(a) to superdiffusion $1 < \nu \leq 2$ in fig. 4(b)-(c) and the again from superdiffusion (subballistic) to hyperdiffusion (superballistic) $2 < \nu$ in fig. 4(d).

Conclusions. —We have seen the two different noise role on diffusion, white noise usually has negative effect on spreading of the wave packet but in some circumstances it can improve the diffusion rate significantly. This results can theoretically simulated in lab and it can have use in cooling nano-wiring instrument because the bigger diffusion rate means better heat conductance. Also the quantum diffusion rate can be extensively tuned by amount of induced white noise. That is we can manually induced noise to a system to drive it to an exact amount of diffusion rate and it can have uses in the nanoelectronic circuits.

Acknowledgment. — We thank Dr. F.Shahbazi and Dr. M. Amini for their helpful comments.

**REFERENCES**

[1] S. Abe and H. Hiramoto, Phys. Rev. A **36**, 5349 (1987).
[2] F. M. Izrailev, T. Kottos, A. Politi, and G. P. Tsironis, Phys. Rev. E **55**, 9711603 (1997).
[3] I. F. Herrera-González, J. a. Méndez-Bermúdez, and F. M. Izrailev, Phys. Rev. E **90**, 042115 (2014).
[4] L. Hufnagel, R. Ketzmerick, T. Kottos, and T. Geisel, Phys. Rev. E **64**, 1 (2001).
[5] J. Fitzsimons and J. Twamley, Phys. Rev. A **72**, 50301 (2005).
[6] Z. Zhang, P. Tong, J. Gong, and B. Li, Phys. Rev. Lett. **108**, 70603 (2012).
[7] S. Stützer, T. Kottos, A. Tünnermann, S. Nolte, D. N. Christodoulides, and A. Szameit, Opt. Lett. **38**, 4675 (2013).
[8] L. Gmachowski, J. Aerosol Sci. **57**, 194 (2013).
[9] Y. Sagi, M. Brook, I. Almog, and N. Davidson, Phys. Rev. Lett. **108**, 3 (2012).
[10] R. Ketzmerick, K. Kruse, S. Kraut, and T. Geisel, Phys. Rev. Lett. **79**, 1959 (1997).
[11] J. Bellissard, Transport **99**, 587 (2000).
[12] L. Levi, Y. Krivolapov, S. Fishman, and M. Segev, Nat. Phys. **8**, 912 (2012).
[13] A. Peruzzo, M. Lobino, J. C. F. Matthews, N. Matsuda, A. Politi, K. Poulios, X.-Q. Zhou, Y. Lahini, N. Ismail, K. Wörhoff, Y. Bromberg, Y. Silberberg, M. G. Thompson, and J. L. OBrien, Science **329**, 1500 (2010).
[14] A. F. Abouraddy, G. Di Giuseppe, D. N. Christodoulides, and B. E. a Saleh, Phys. Rev. A - At. Mol. Opt. Phys. **86**, 1 (2012).
[15] G. W. Ford and R. F. O'Connell, Am. J. Phys. **70**, 319 (2002).
[16] H. P. Breuer, Phys. Rev. A - At. Mol. Opt. Phys. **75**, 1 (2007).
[17] a. Shabani, J. Roden, and K. B. Whaley, Phys. Rev. Lett. **112**, 23 (2014).
[18] C. Gardiner and P. Zoller, *The Quantum World of Ultra-Cold Atoms and Light: Foundations of Quantum Optics* (Imperial College Press, 2014).
[19] C. W. Gardiner and P. Zoller, *Quantum Noise : A Handbook of Markovian and Non-Markovian Quantum Stochastic Methods with Applications to Quantum Optics* (2004).
[20] Y. Hu, Z. Cai, M. A. Baranov, and P. Zoller, Phys. Rev. B **92**, 165118 (2015).
[21] H. Haken and G. Strobl, Z. Phys. **262**, 135 (1973).
[22] L. Novo, M. Mohseni, and Y. Omar, Sci. Rep. **6**, 18142 (2016).
[23] P. G. Harper, Proc. Phys. Soc. Sect. A **68**, 879 (1955).
[24] G. Roati, C. D'Errico, L. Fallani, M. Fattori, C. Fort, M. Zaccanti, G. Modugno, M. Modugno, and M. Inguscio, Nature **453**, 895 (2008).